\documentclass[conference]{IEEEtran} 
\usepackage[noadjust]{cite}
\usepackage{amsmath, amsfonts, amssymb}
\usepackage{algorithmic}
\usepackage{graphicx}
\usepackage{textcomp}
\usepackage{xcolor}
\usepackage{caption}
\usepackage{subcaption}
\usepackage{algorithm}
\usepackage{array}
\usepackage{stfloats}
\usepackage{url}
\usepackage{verbatim}
\usepackage{pifont}
\usepackage[normalem]{ulem}
\usepackage{makecell}
\usepackage{array}
\usepackage{multirow}
\usepackage{multicol}
\usepackage[nolist,nohyperlinks]{acronym}

\acrodef{OFDM}[OFDM]{Orthogonal Frequency Division Multiplexing}
\acrodef{FFT}[FFT]{Fast Fourier Transform}
\acrodef{IFFT}[IFFT]{Inverse Fast Fourier Transform}
\acrodef{FLOPs}[FLOPs]{floating point operations}
\acrodef{NFLOPs}[NFLOPs]{number of floating point operations}
\acrodef{NeuralRx}[NeuralRx]{neural receiver}
\acrodef{ResNet}[ResNet]{residual network}
\acrodef{ResNet-T}[ResNet-T]{traditional residual network}
\acrodef{ResNet-SS}[ResNet-SS]{residual network-split and shuffle}
\acrodef{GELU}[GELU]{Gaussian error linear unit}
\acrodef{MAC}[MAC]{memory access cost}
\acrodef{CNN}[CNN]{convolutional neural network}
\acrodef{TTI}[TTI]{transmission time interval}
\acrodef{LS}[LS]{least squares}
\acrodef{LMMSE}[LMMSE]{linear minimum mean square error}
\acrodef{CSI}[CSI]{channel state information}
\acrodef{RNN}[RNN]{recurrent neural network}
\acrodef{LLR}[LLR]{log-likelihood ratio}
\acrodef{DS-Conv-2D}[DS-Conv-2D]{two depthwise separable 2D convolution}
\acrodef{Conv-2D}[Conv-2D]{2D convolution}
\acrodef{ReLU}[ReLU]{rectifier linear unit}
\acrodef{LDPC}[LDPC]{low-density parity check}
\acrodef{QAM}[QAM]{quadrature amplitude modulation}
\acrodef{PRB}[PRB]{physical resource block}
\acrodef{DMRS}[DMRS]{demodulation reference signal}
\acrodef{CP}[CP]{cyclic prefix}
\acrodef{TDL}[TDL]{tapped delay line}
\acrodef{CDL}[CDL]{clustered delay line}
\acrodef{AWGN}[AWGN]{additive white Gaussian noise}
\acrodef{PDSCH}[PDSCH]{physical downlink shared channel}
\acrodef{PUSCH}[PUSCH]{physical uplink shared channel}
\acrodef{5G-NR}[5G-NR]{5G New Radio}
\acrodef{C2R}[$\mathbb{C}2\mathbb{R}$]{complex to real}
\acrodef{NL}[NL]{normalization layer}
\acrodef{NParam}[NParam]{number of parameters}
\acrodef{BCE}[BCE]{binary cross entropy}
\acrodef{SNR}[SNR]{signal-to-noise-ratio}
\acrodef{RMS}[RMS]{root mean square}
\acrodef{BLER}[BLER]{block error rate}

\newcommand{\tb}{\textbf}

\definecolor{dark_green}{rgb}{0, 0.5, 0}
\definecolor{dark_yellow}{rgb}{0.75, 0.75, 0}

\begin{document}

\title{Low-Complexity OFDM Deep Neural Receivers}

\author{Ankit Gupta, Onur Dizdar, Yun Chen, Fehmi Emre Kadan, Ata Sattarzadeh, and Stephen Wang.\\
VIAVI Marconi Labs, VIAVI Solutions Inc., Stevenage SG1 2AN, UK.\\
Email: {ankit.gupta, onur.dizdar, yun.chen, fehmiemre.kadan, ata.sattarzadeh, stephen.wang}@viavisolutions.com}



\maketitle

\begin{abstract}
Deep neural receivers (NeuralRxs) for Orthogonal Frequency Division Multiplexing (OFDM) signals are proposed for enhanced decoding performance compared to their signal-processing based counterparts. However, the existing architectures ignore the required number of epochs for training convergence and floating-point operations (FLOPs), which increase significantly with improving performance. 
To tackle these challenges, we propose a new residual network (ResNet) block design for OFDM NeuralRx. Specifically, we leverage small kernel sizes and dilation rates to lower the number of FLOPs (NFLOPs) and uniform channel sizes to reduce the memory access cost (MAC). The ResNet block is designed with novel channel split and shuffle blocks, element-wise additions are removed, with Gaussian error linear unit (GELU) activations. Extensive simulations show that our proposed NeuralRx reduces NFLOPs and improves training convergence while improving the decoding accuracy. 
\end{abstract}

\section{Introduction}
\IEEEPARstart{D}{eep} \acp{NeuralRx} substitute a neural network for several signal processing blocks, including channel estimation, interpolation, equalization, and symbol de-mapping. By removing design assumptions inherent in signal processing blocks and processing entire \ac{TTI} together, the joint design enhances the decoding performance compared to \ac{LMMSE} receivers that rival even 5G-NR receivers with full \ac{CSI} knowledge~\cite{Honkala2021, Faycal2022, Mei2024, Korpi2023, Fischer2022, Gupta2023, Pihlajasalo2023, Raviv2023, Xie2024, gupta2024spikingrx}.

Among various architectures, such as \acp{RNN} that focus on the sequential nature of the \ac{OFDM} grid over time and frequency~\cite{Fischer2022}, the \acp{CNN} have gained prominence for their ability to capture spatial dependencies with lesser parameters. The \acp{NeuralRx} process the OFDM grid through multiple \ac{ResNet} blocks to generate \acp{LLR}. A similar \ac{ResNet} block architecture is utilized across implementations with varying number of blocks and hyper-parameters. These ``\ac{ResNet-T}" blocks are designed with \ac{DS-Conv-2D} layers from MobileNet~\cite{MobileNet2017}, that achieves similar performance to \ac{Conv-2D}-based designs while improving computation efficiency. The \ac{ResNet-T} block utilizes \ac{ReLU} activations that remains computationally efficient and simple, however, its hard thresholding of all the negative values limits its learning capabilities by discarding the informative negative values.

While decoding performance is critical, practical deployment also demands efficient computation. The complexity of \ac{NeuralRx} can be evaluated using the number of parameters that measures the memory footprint however, it doesn’t necessarily correlate with runtime efficiency. On the other hand, the \ac{NFLOPs} provides a theoretical measure of hardware-independent workload estimate. Prior \ac{ResNet-T}-based \ac{NeuralRx} works~\cite{Honkala2021, Fischer2022} analyze the number of parameters for complexity evaluation. In~\cite{Honkala2021}, authors vary hyper-parameters, such as the number of \ac{ResNet} blocks, filters, kernels, the size of depth multiplier, and dilation rate to design the optimal architecture. The authors propose adaptive learning-based training for designing low-complexity \ac{NeuralRx} in \cite{Fischer2022}. Although these previous works utilize \ac{ResNet-T} blocks, a balanced trade-off between performance and computational efficiency still remains a challenge. Further, none of the aforementioned works focus on training convergence speed, essential for online re-training scenarios. 

In this work, we design an energy-efficient \ac{NeuralRx} with a novel \ac{ResNet} block, named \textit{\ac{ResNet-SS}}. The proposed \ac{ResNet-SS} combines the split and shuffle operations in ShuffleNet~\cite{ShuffleNet2017} with \ac{ResNet-T} blocks. ResNet-SS splits channels to reduce the \ac{NFLOPs} and \ac{MAC} by eliminating element-wise additions compared to \ac{ResNet-T}. Additionally, channel shuffle is utilized to act as stochastic regularization and \ac{GELU} activation~\cite{GELU23} provides smoother and more informative gradient flow than \ac{ReLU}. We design a low-complexity \ac{NeuralRx} architecture as a combination of \ac{ResNet-T} and \ac{ResNet-SS} blocks. Further, we propose utilizing small $3\times 3$ kernel sizes and dilation rates $=1$ to lower \ac{NFLOPs} and uniform channel sizes to reduce \ac{MAC}. We analyze training convergence, decoding performance, and computational complexity of proposed NeuralRx and compare it with traditional NeuralRx designs to demonstrate its advantages.

\section{System Model} \label{sec:system_model}
\begin{figure}[t!]
    \centering
    \includegraphics[scale=0.425]{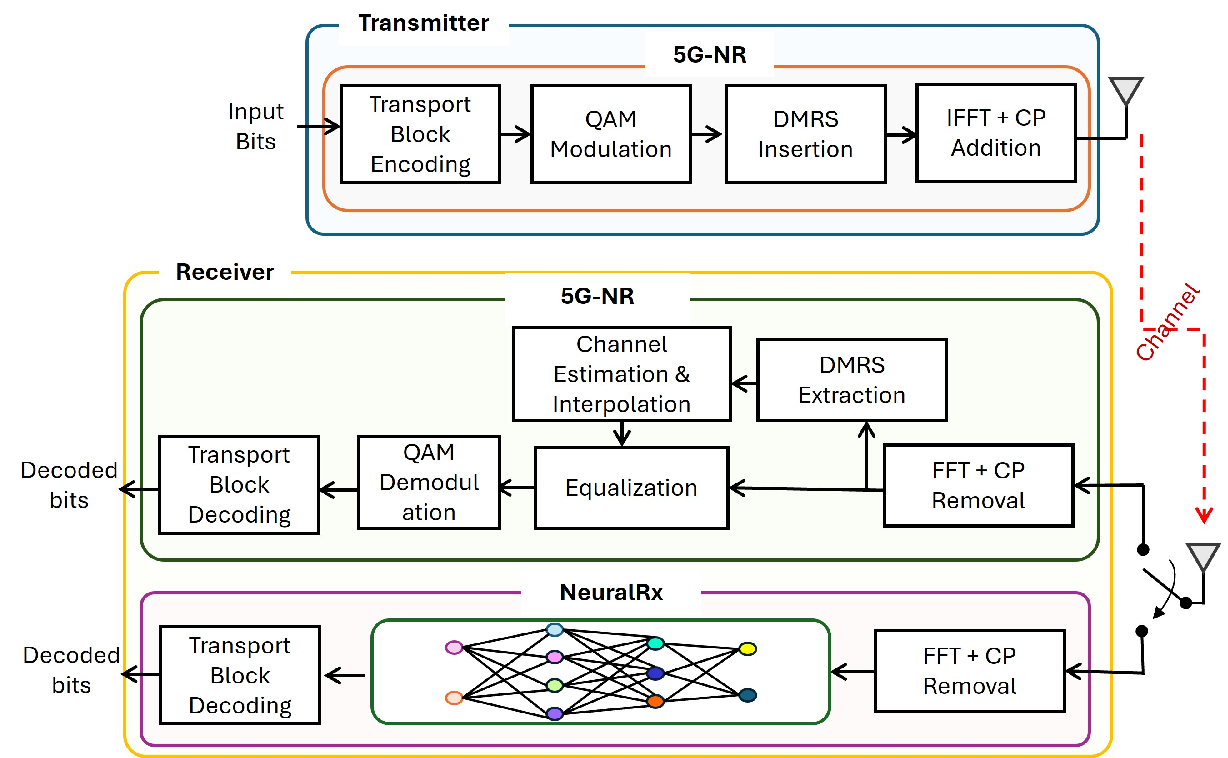}
    \caption{Block diagram of 5G-NR transceiver and \ac{NeuralRx}.}
   \label{fig:system_model}
\end{figure}

As shown in Fig.~\ref{fig:system_model}, we consider a transmitter with $N_T$ antennas transmitting single layer data over \ac{5G-NR} \ac{PDSCH}/\ac{PUSCH} to receiver with $N_R$ antennas.

At the transmitter, the information bits are fed into the transport block encoder, that performs \ac{LDPC} coding, rate-matching, scrambling, to output a series of codewords. These are converted to complex baseband symbols by \ac{QAM} and mapped to \acp{PRB} in a resource grid. Then, \ac{DMRS} symbols are inserted to specific \ac{OFDM} symbols by occupying all resource elements. The resulting resource grid is passed through an \ac{IFFT} block and \ac{CP} is added to every \ac{OFDM} symbol. The obtained signal propagates through the 3GPP TR38.901 \ac{TDL}/\ac{CDL} channels and distorted by \ac{AWGN}. The receiver removes the \ac{CP} and applies a \ac{FFT} to obtain,
\begin{align}
\mathbf{y}_{m,n} = \mathbf{H}_{0,m,n} \tb{w}_{m,n} x_{m,n} + \mathbf{z}_{m,n} = \mathbf{h}_{m,n}x_{m,n} + \mathbf{z}_{m,n},
\label{eqn:received}
\end{align}
where $x_{m,n}\in\mathbb{C}$ and $\mathbf{y}_{m,n}\in\mathbb{C}^{N_R\times 1}$ denote the transmitted and received signals, respectively, $\mathbf{H}_{0,m,n}\in\mathbb{C}^{N_R\times N_T}$ denotes the channel between the transmitter and receiver, $\tb{w}_{m,n} \in \mathbb{C}^{N_T \times 1}$ is the precoder/beamformer designed by the transmitter to transmit the single layer data $x_{m,n}$, $\mathbf{h}_{m,n}=\tb{H}_{0,m,n}\tb{W}_{m,n} \in \mathbb{C}^{N_R \times 1}$ is the effective channel between transmitter and receiver, and $\mathbf{z}_{m,n} \sim \mathcal{CN}(\tb{0},N_0\tb{I}) \in\mathbb{C}^{N_R\times 1}$ is the \ac{AWGN} at the $m$-th \ac{OFDM} symbol and $n$-th subcarrier for $m\in\left\{0,\ \ldots,\ \ M-1\right\}$ and $n\in\left\{0,\ldots,\ \ N-1\right\}$.

In a traditional \ac{5G-NR} receiver, channel estimation and interpolation are performed to obtain $\mathbf{\hat{H}}\in\ \mathbb{C}^{N_R\times M\times N}$ after \ac{FFT} by utilizing the known \ac{DMRS} symbols denoted by by $\mathbf{P}\in\mathbb{C}^{N_R\times M\times N}$.
Note that \ac{DMRS} is also precoded/beamformed as data symbols and hence the effective channel is valid for \ac{DMRS} as well. 
The equalized symbols are passed to a signal demapper to obtain the \ac{LLR} values, which are passed to the transport block decoder to get the decoded bits.

\section{Proposed \ac{OFDM} \ac{NeuralRx} Design}
In this section, we describe the design and training of \ac{NeuralRx} with the proposed \ac{ResNet} blocks.

\subsection{Preliminaries on NeuralRx Design}
\begin{figure*}[t!]
    \centering
    \begin{subfigure}{0.28\linewidth}
        \centering
        \includegraphics[scale=0.72]{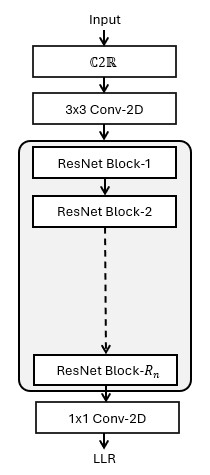}
        \caption{NeuralRx architecture.}
        \label{fig:neuralrx_arch}
    \end{subfigure}
    \begin{subfigure}{0.28\linewidth}
        \centering
        \includegraphics[scale=0.75]{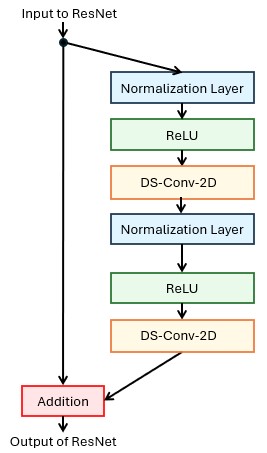}
        \caption{Traditional ResNet-T block.} 
        \label{fig:trad_resnet}
    \end{subfigure}
    \begin{subfigure}{0.28\linewidth}
        \centering
        \includegraphics[scale=0.57]{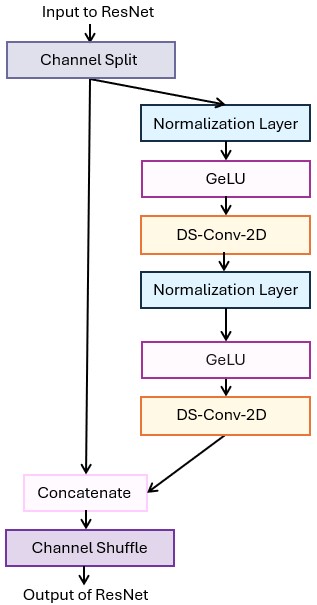}
        \caption{Proposed ResNet-SS block.}
        \label{fig:resnet_b}
    \end{subfigure}
    \caption{Architecture of NeuralRx with ResNet blocks.}
    \label{fig:resnet_blocks}
\end{figure*}

We depict the typical \ac{NeuralRx} architecture in Fig.~\ref{fig:neuralrx_arch} comprising of input layer, \ac{C2R} layer, \ac{Conv-2D} layer, multiple \ac{ResNet-T} blocks and \ac{Conv-2D} layer with Sigmoid activation to produce the \ac{LLR} outputs. In Fig.~\ref{fig:trad_resnet}, we show a \ac{ResNet-T} block, wherein the input is passed through \ac{NL}, \ac{ReLU} activation and \ac{DS-Conv-2D} layers, before being added to the input by the skip connection. Conventionally, the complexity of ResNet-T is calculated in terms of \ac{NParam}, and the ratio of complexities in ResNet-T with and without \ac{DS-Conv-2D} layers is calculated as~\cite{Honkala2021, Faycal2022, Mei2024, Korpi2023, Fischer2022, Gupta2023, Pihlajasalo2023, Raviv2023, Xie2024, gupta2024spikingrx}
\begin{align}
    \dfrac{\text{NParam}_{\text{DS-Conv-2D}}}{\text{NParam}_{\text{Conv-2D}}} \!=\!\frac{D_m C_{\text{in}} (K_l^2\!+\! C_{\text{out}})}{C_{\text{in}} C_{\text{out}} K_l^2} \!=\!\frac{D_m}{C_{\text{out}}}\!+\!\frac{D_m}{K_l^2} \label{eq:params}
\end{align}
where $D_m$ is the depth multiplier for \ac{DS-Conv-2D} layer, $C_{\text{in}}, C_{\text{out}}$ denote the input and output filter sizes, respectively, and $K_l$ denotes the kernel size. For example, when $D_m=1$ and $K_l = 3$, the design with \ac{DS-Conv-2D} layer reduces the number of parameters by approximately $9$ times compared to one with \ac{Conv-2D} with similar performance. Thus, \ac{DS-Conv-2D} remains popular for \ac{NeuralRx} designs~\cite{Honkala2021, Faycal2022, Mei2024, Korpi2023} and also adopted it in this work.

\subsection{Proposed Efficient \ac{ResNet} Blocks}
\label{sec:proposedresnet}
\subsubsection{Kernel Size and Dilation Rate}  
Prior works assess complexity using the NParam as in \eqref{eq:params}, that may not reflect runtime efficiency. Instead, we propose to utilize \ac{NFLOPs} as:
\begin{align}
\frac{\text{NFLOPs}_{\text{DS-\!Conv-2D}}}{\text{NFLOPs}_{\text{Conv-2D}}}\!\!=\!\!\frac{D_m\!H W C_{\text{in}} (\!K_{\text{eff}}^2\!+\!C_{\text{out}}\!)}{H W C_{\text{in}} C_{\text{out}} K_{\text{eff}}^2}\!\!=\!\!\frac{D_m}{C_{\text{out}}}\!\!+\!\!\frac{D_m}{K_{\text{eff}}^2} \label{eq:flops}
\end{align}
where $H, W$ denote the height and width of the input, respectively, and $K_{\text{eff}}$ denotes the effective kernel size based on the kernel size $K_l$ and dilation rate $D_l$, given as
\begin{align}
    K_{\text{eff}} = D_l(K_l-1)+1. \label{eq:kernel_size}
\end{align}
Unlike the number of parameters in \eqref{eq:params}, \ac{NFLOPs} also considers height and width of input. Further, dilation rate $D_l$ has no impact on parameter count because it only spreads out the weights over larger receptive field, it impacts \ac{NFLOPs}. Thus, prior \ac{NeuralRx} works were able to consider large dilation rates, {\sl i.e.,} $D_l\gg 1$, without realizing the dramatic increase in \ac{NFLOPs}. As an example, consider the case for $K_l=3$. Then \ac{NFLOPs} for $D_l=\{3, 5\}$ increases by $\approx 6$ and $\approx 14$ times compared to $D_l = 1$. Moreover, the \ac{MAC} also increases with dilation rate due to larger memory gaps and inefficient cache usage, further impacting the runtime efficiency. 

Since $\text{\ac{NFLOPs}}\propto K_{\text{eff}}^2$, we propose to design \ac{ResNet} blocks with the lowest kernel size $K_l=3$ and dilation rate $D_l=1$ to minimize both \ac{NFLOPs} and \ac{MAC}. To compensate for reduced receptive field, we can increase the number of \ac{ResNet} blocks to maintain decoding performance. This increases \ac{NFLOPs} linearly but keep it significantly lower than that when $D_l\gg 1$.

\subsubsection{Filter Size} The \ac{MAC} has a lower bound given by \ac{NFLOPs} for \ac{DS-Conv-2D} layer, satisfying~\cite{ShuffleNet2017}
\begin{align}
    \text{MAC}\!\geq\!2\sqrt{HW\text{NFLOPs}_{\text{DS-Conv-2D}}} + \dfrac{\text{NFLOPs}_{\text{DS-Conv-2D}}}{HW}
\end{align}
where the lower bound is achieved for equal input and output filter sizes, i.e., $C_{\text{in}}=C_{\text{out}}$. Based on this observation, we propose using same channel size $C=C_{\text{in}}=C_{\text{out}}$ for all \ac{DS-Conv-2D} and \ac{Conv-2D} layers in \ac{NeuralRx}. Indeed, several existing works showed good performance with same channel size, but their focus was not on \ac{MAC} optimization~\cite{Faycal2022}.

\begin{algorithm}[t]
\caption{Operations in ResNet-SS Block}\label{algo_2}
\begin{algorithmic}[1]
    \STATE \textbf{Input:} Signal $\mathbf{X} \in \mathbb{R}^{H \times W \times C}$, channel split $C' = C/2$ and Groups $G = C/2$.
    \STATE Split $\mathbf{X}$ into two branches:
    \begin{equation}
        \mathbf{X} = [\mathbf{X}'\in \mathbb{R}^{H \times W \times C'}, \mathbf{X}''\in \mathbb{R}^{H \times W \times (C - C')}]\vspace{-0.15cm}
    \end{equation}
    \STATE Process the $\mathbf{X}''$ branch:
    \begin{align}
        \mathbf{X}''_{\text{Processed}} &= \text{DS-Conv-2D}(\text{GeLU}(\text{NL}(\nonumber\\
        & \qquad \text{DS-Conv-2D}(\text{GeLU}(\text{NL}(\mathbf{X}''))))))
    \end{align}
    \STATE Concatenate two branches:
    \begin{equation}
        \mathbf{X}_{\text{concat}} = \text{Concat}(\mathbf{X}', \mathbf{X}''_{\text{Processed}})
    \end{equation}
    \STATE Reshape $\mathbf{X}_{\text{concat}}$ to $\mathbf{X}_{\text{reshaped}}$:
\begin{equation}
    \mathbf{X}_{\text{reshaped}} = \text{Reshape}(\mathbf{X}_{\text{concat}}, (H, W, G, {C}\big/{G}))
\end{equation}
\STATE Transpose groups and channels in $\mathbf{X}_{\text{concat}}$ to $\mathbf{X}_{\text{transposed}}$:
\begin{equation}
    \mathbf{X}_{\text{transposed}} = \text{Transpose}(\mathbf{X}_{\text{reshaped}}, (0,1,3,2))
\end{equation}
\STATE Reshape $\mathbf{X}_{\text{transposed}}$ to original shape $\mathbf{X}_{\text{shuffled}}$:
\begin{equation}
    \mathbf{X}_{\text{shuffled}} = \text{Reshape}(\mathbf{X}_{\text{transposed}}, (H, W, C))
\end{equation}
\end{algorithmic}
\end{algorithm}

\subsubsection{\ac{ResNet-SS} Block Design}
Small mobile devices with limited \ac{NFLOPs} budget can only support limited filter sizes because the \ac{NFLOPs} in \ac{DS-Conv-2D} or \ac{Conv-2D} layers scales quadratically by filter sizes, i.e, $\text{NFLOPs}\propto C^2$ in \eqref{eq:flops}. In order to avoid such complexity increase, we propose incorporating channel split and shuffle operations in the \ac{ResNet-T} blocks to reduce \ac{NFLOPs}, based on the ShuffleNet-v2 design~\cite{ShuffleNet2017}\footnote{We note directly using Shufflenet-v2 blocks lead to degraded performance.}
As shown Fig.~\ref{fig:resnet_b}, we refer to these novel \ac{ResNet} blocks as \ac{ResNet-SS}. As detailed in Algorithm~\ref{algo_2}, the channel split operation splits the input channels into two groups each with $(C'=C/2)$ filter size and processes only one group, reducing \ac{NFLOPs} by a factor of $4$ times in \ac{DS-Conv-2D} or \ac{Conv-2D} layers ($\text{NFLOPs}\propto (C')^2\triangleq(C/2)^2$). Furthermore, the element-wise addition in traditional \ac{ResNet-T} block requires $H\times W\times C$ \ac{FLOPs} and increases \ac{MAC}. Thus, we replace this addition with channel concatenation, which requires no \ac{FLOPs} and reduces \ac{MAC}, while preserving the feature richness of the original input with filter size $C$. To mitigate potential performance degradation due to processing with fewer convolutional channels, we incorporate the channel shuffle that interleaves the channels from both chains, ensuring the information exchange and acting as an inherent stochastic regularization. Unlike other regularization techniques such as dropout and $L2$-regularization that requires $2$ \ac{FLOPs} per neuron, the proposed channel shuffle operation requires no additional \ac{FLOPs}, as it solely consists of tensor memory manipulations. We can further reduce the \ac{MAC} by combining the channel split, concatenate, and channel shuffle operations into a single element-wise operation~\cite{ShuffleNet2017}. We propose to utilize the \ac{GELU} activation~\cite{GELU23} over \ac{ReLU}, as
\begin{align}
    \text{GELU}(x)\!\approx\!0.5x\!\left[ 1\!+\!\text{tanh}\left(\!\sqrt{2/\pi} \left(x+0.044715x^3\right)\!\right)\!\right]
\end{align}
Unlike \ac{ReLU}, \ac{GELU} remains differentiable everywhere, mimicking dropout-like behavior for negative inputs, acts as a soft gate for values near $0$, and avoids dead neuron problem. 


\subsection{Proposed \ac{NeuralRx} Architecture and Training}
\begin{table}[!t]
\vspace{0.1cm}
\centering
\caption{Architecture of combined ResNet-T-SS block.} \label{tab:arch_resnet_t_ss}
\begin{tabular}{c|c|c|c|c}
\hline
\textbf{Layers} & \textbf{Repeat} & \textbf{Filter Size} & \textbf{Kernel Size} & \textbf{Dilation Rate} \\ \hline
ResNet-SS    & $2$  & $128$   &  $(3, 3)$      & $(1, 1)$      \\ \hline
ResNet-T    & $1$  & $128$   &  $(3, 3)$      & $(1, 1)$      \\ \hline
\end{tabular}
\end{table}

\begin{table}[!t]
\centering
\caption{Architecture for Proposed \ac{NeuralRx}.} \label{tab:arch_resnet_ss}
\begin{tabular}{c|c|c|c|c}
\hline
\textbf{Layers} & \textbf{Repeat} & \textbf{Filter Size} & \textbf{Kernel Size} & \textbf{Dilation Rate} \\ \hline
Conv-2D      & $1$ & $128$      &  $(3, 3)$      & $(1, 1)$      \\ \hline
ResNet-T-SS    & $\lfloor R_n/2 \rfloor$  & $128$   &  $(3, 3)$      & $(1, 1)$      \\ \hline
ResNet-SS    & $2$  & $128$   &  $(3, 3)$      & $(1, 1)$      \\ \hline
Conv-2D      & $1$  & $128$      &  $(1, 1)$      & $(1, 1)$      \\ \hline
\end{tabular}
\end{table}

\subsubsection{Proposed \ac{NeuralRx} Architecture}\label{sec:nrx_arch}
We keep the \ac{NeuralRx} architecture as shown in Fig.~\ref{fig:neuralrx_arch} with $R_n$ \ac{ResNet} blocks. Based on the discussions in Section~\ref{sec:proposedresnet}, we design the \ac{NeuralRx} architecture with equal filter sizes $C=128$, smaller kernel size $(3, 3)$ and dilation rate $(1, 1)$. By simulations, we find that replacing all \ac{ResNet-T} blocks with the \ac{ResNet-SS} block results in suboptimal performance due to limited filter sizes. However, alternating between proposed $2$ \ac{ResNet-SS} and $1$ \ac{ResNet-T} block provides balance between performance, complexity and training convergence. For clarity, we define this configuration as ResNet-T-SS block in Table~\ref{tab:arch_resnet_t_ss}. Finally, Table~\ref{tab:arch_resnet_ss} summarizes the \ac{NeuralRx} architecture. The proposed ResNet-T-SS blocks can be used in other state-of-art \ac{NeuralRx} with \ac{ResNet-T} blocks with their configurations for improved convergence, performance, and reduced complexity. 

\subsubsection{\ac{NeuralRx} Training and Evaluation}
\begin{table}[t]
\vspace{0.1cm}
\caption{Parameters for Training and Testing.}
\label{tab:train_test_params}
\centering 
\begin{tabular}{m{2.2cm}|m{2cm}|m{1.8cm}|m{1cm}} 
\hline
\textbf{Parameter} & \textbf{Training/ Validation} & \textbf{Testing} & \textbf{Random- ization} \\ [0.5ex] 
\hline
\hline
Rx/Tx Antennas & \multicolumn{2}{c|}{$N_R = 2, N_T = 1$} & None\\
\hline
Carrier Freq. & \multicolumn{2}{c|}{$4$ GHz} & None\\
\hline
Numerology & \multicolumn{2}{c|}{$1$ ($30$ kHz subcarrier spacing)} & None\\
\hline
Number of \acp{PRB} & \multicolumn{2}{c|}{$21.33$ ($256$ subcarriers)} & None\\
\hline
Symbol Duration & \multicolumn{2}{c|}{$38.02\;\mu$s} & None\\
\hline
CP Duration & \multicolumn{2}{c|}{$4.68\;\mu$s} & None\\
\hline
\ac{TTI} Length & \multicolumn{2}{c|}{$14$ OFDM Symbols ($1$ ms)} & None\\
\hline
Modulation & \multicolumn{2}{c|}{64-\ac{QAM} ($M_O=6$)} & None\\
\hline
Code-rate & \multicolumn{2}{c|}{0.5} & None\\
\hline
\ac{DMRS}  & \multicolumn{2}{c|}{$1$ or $2$ OFDM symbols per slot} & None\\
\hline
Channel Model & \ac{CDL}-A, C, E, \ac{TDL}-A, C, E & \ac{CDL}-B, D, \ac{TDL}-B, D & Uniform\\
\hline
$E_b/N_0$ & \multicolumn{2}{c|}{$0-35$ dB} & Uniform\\
\hline
RMS Delay Spread & $10-1000$ ns  & $100-700$ ns & Uniform\\
\hline
Doppler Shift & \multicolumn{2}{c|}{$0-700$ Hz} & Uniform\\
\hline
\end{tabular}
\end{table}
The \ac{NeuralRx} solves a multi-label binary classification problem by minimizing the \ac{BCE} loss between the transmitted bits and \acp{LLR} obtained from the \ac{NeuralRx}. We utilize the AdamW optimizer with learning-rate $\eta=10^{-3}$, batch size $B=128$ and $\approx 10.25$ million \acp{TTI} for training.

\subsubsection{Generalizability}
The proposed \ac{NeuralRx} considers 3GPP TR38.901 antenna pattern with vertical polarization and remains generalizable to varying \ac{DMRS} configurations, 3GPP TR38.901 \ac{TDL}/\ac{CDL} channel profiles, \ac{SNR} (measured in terms of energy per bit to noise power spectral density ratio, $E_b/N_0$), Doppler spread and \ac{RMS} delay spread. We summarize the training and testing parameters in Table~\ref{tab:train_test_params}.

\section{Performance Evaluation}\label{sec:performance}
In this section, we evaluate the performance of the proposed and baseline receivers served by traditional \ac{5G-NR} transmitter. 

\subsection{Baseline Methods} 
We evaluate three \ac{5G-NR} baseline receivers: (1) \textit{PCSI-Rx}, assuming perfect \ac{CSI} for upper-bound performance, (2) \textit{LS-Rx}, utilizing \ac{LS} channel estimation with low-complexity linear interpolation for lower-bound performance, and (3) \textit{LMMSE-Rx}, employing high-complexity \ac{LMMSE} channel estimation for improved performance over LS-Rx. 

\begin{table}[!t]
\centering
\caption{Architecture for \ac{NeuralRx} with \ac{ResNet-T}.} \label{tab:arch_resnet_t}
\begin{tabular}{c|c|c|c|c}
\hline
\textbf{Layers} & \textbf{Repeat} & \textbf{filter sizes} & \textbf{Kernel Size} & \textbf{Dilation Rate} \\ \hline
Conv-2D      & $1$ & $128$      &  $(3, 3)$      & $(1, 1)$      \\ \hline
ResNet-T  & $R_n$  & $128$   &  $(3, 3)$      & $(1, 1)$      \\ \hline
Conv-2D    & $1$  & $128$      &  $(1, 1)$      & $(1, 1)$      \\ \hline
\end{tabular}
\end{table}

\begin{figure*}[t!]
    \centering
    \includegraphics[scale=0.5]{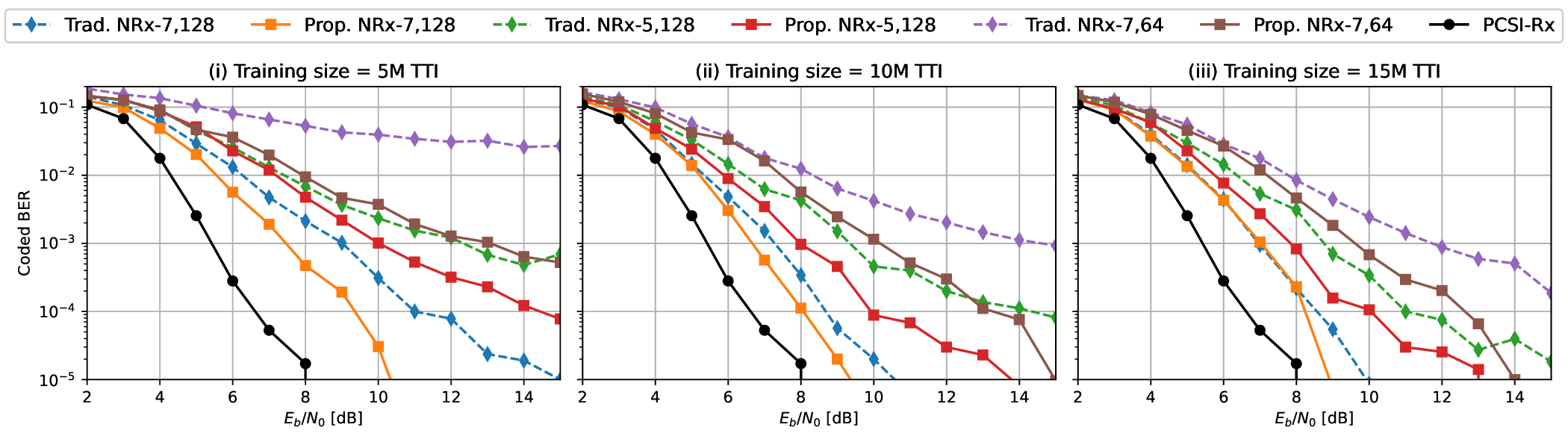}
    \caption{Training convergence for Y-NeuralRx (parameters in Table~\ref{tab:arch_resnet_t}) and proposed NeuralRx (parameters in Table~\ref{tab:arch_resnet_ss}).}
   \label{fig:train_converg}
\end{figure*}

We evaluate two baseline \acp{NeuralRx} classified based on the inputs as: (i) Y-NeuralRx~\cite{Faycal2022}, that only receives frequency-domain OFDM resource grid $\mathbf{Y}\in\mathbb{C}^{N_R\times M\times N}$ in \eqref{eqn:received} as input, and (ii) YHP-NeuralRx~\cite{Honkala2021}, that employs the \ac{DMRS} extraction and \ac{LS} channel estimation blocks of \ac{5G-NR} receiver chain with the \ac{NeuralRx}, thereby receiving OFDM resource grid $\mathbf{Y}\in\mathbb{C}^{N_R\times M\times N}$ in \eqref{eqn:received}, \ac{LS} channel estimate $\mathbf{\hat{H}}\in\ \mathbb{C}^{N_R\times M\times N}$, and \ac{DMRS} symbols $\mathbf{P}\in\mathbb{C}^{N_R\times M\times N}$ as inputs. Both YHP-NeuralRx~\cite{Honkala2021} and Y-NeuralRx~\cite{Faycal2022} originally utilized \ac{ResNet-T} blocks with varying kernel sizes and larger dilation rates, leading to significantly higher \ac{NFLOPs}, and unequal filter sizes~\cite{Honkala2021}. To isolate the impact of \ac{ResNet} block design, we adopt the YHP-NeuralRx and Y-NeuralRx architectures with \ac{ResNet-T} block from Sec.~\ref{sec:nrx_arch}, referred to as (4) \textit{traditional YHP-NeuralRx} and (5) \textit{traditional Y-NeuralRx}, as given in Table~\ref{tab:arch_resnet_t}. 

\subsection{Training Convergence}

In Fig.~\ref{fig:train_converg}, we evaluate the training convergence for varying \ac{NeuralRx} architectures by training Y-NeuralRx and testing for single \ac{DMRS} per slot and speed of $1-50$ m/s at different training dataset sizes $5, 10, 15$ million \acp{TTI}. We consider \ac{NeuralRx} with varying number of \ac{ResNet} blocks \{5,7\} and filter sizes \{64, 128\}. PCSI-Rx shows the upper bound. Proposed \ac{NeuralRx} (Table~\ref{tab:arch_resnet_ss}) with \ac{ResNet-SS} blocks improve performance up to $5$~dB over traditional \ac{NeuralRx} (Table~\ref{tab:arch_resnet_t}) with only\ac{ResNet-T} blocks for smaller training dataset. As the training dataset size increases the performance gains reduces to $2$~dB. Notably, for resource constrained scenarios, such as \ac{NeuralRx} at battery-powered mobile devices, using smaller \ac{NeuralRx} architectures, \ac{ResNet-SS} blocks provides considerable gains  irrespective of training dataset sizes. On the other hand, for scenarios where there are enough resources and using larger architectures is feasible, such as using \ac{NeuralRx} at base station, the gains are observed for smaller datasets and only at higher \ac{SNR} regime for larger datasets. 
This is because larger architectures have sufficient capacity to learn properly even with the \ac{ResNet-T} block if enough training dataset is available.

\subsection{Decoding Performance Comparison}
\begin{figure*}[t!]
    \centering
    \includegraphics[scale=0.5]{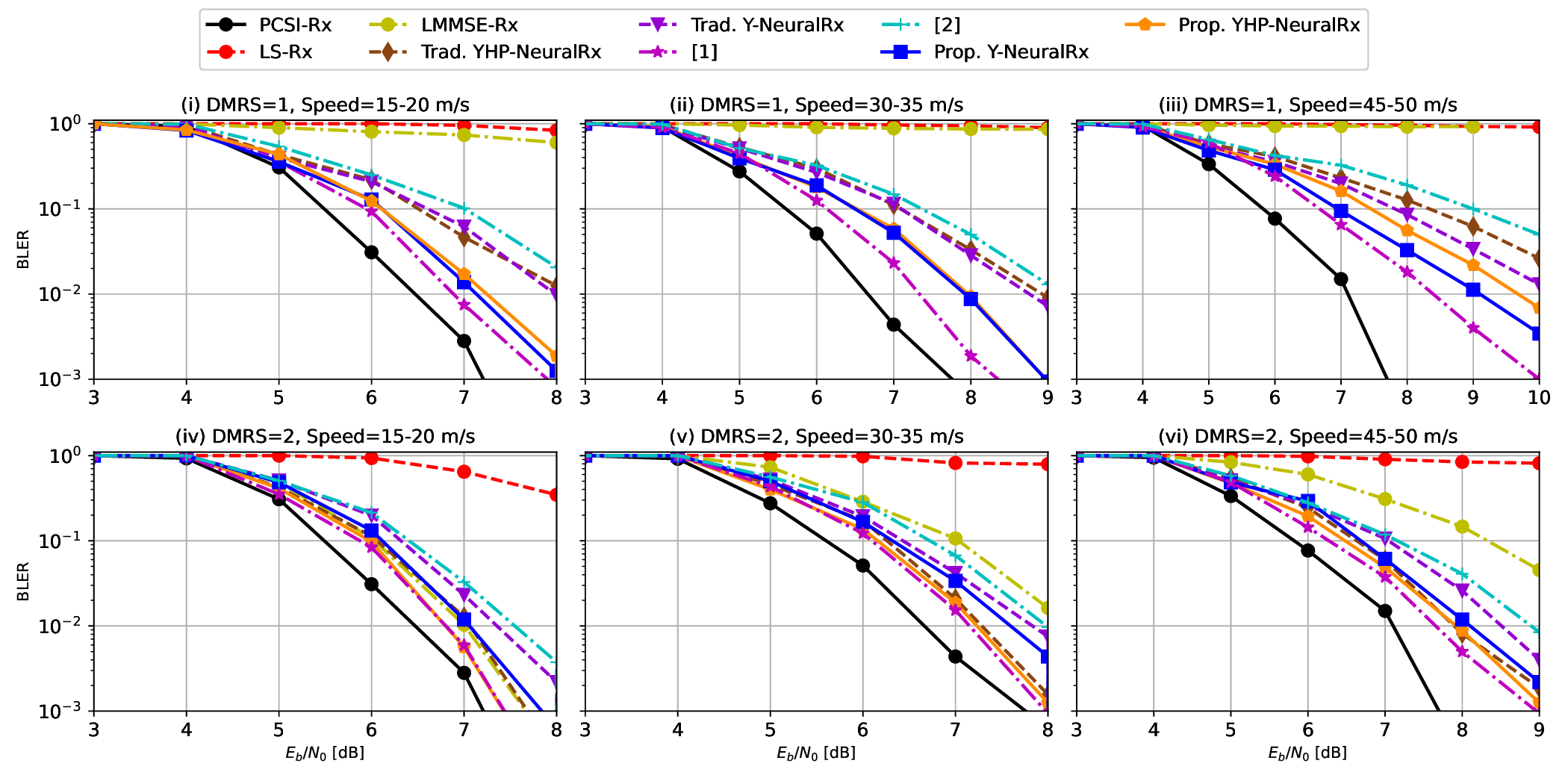}
    \caption{Decoding performance of traditional \ac{NeuralRx} (parameters in Table~\ref{tab:arch_resnet_t}) and proposed \ac{NeuralRx} (parameters in Table~\ref{tab:arch_resnet_ss}).}
   \label{fig:main_perf_evaluation}
\end{figure*}
In Fig.~\ref{fig:main_perf_evaluation}, we compare the decoding performance for baseline and proposed methods. For all scenarios, the \ac{5G-NR} LS-Rx performs the worst due to worst channel estimation and the PCSI-Rx performs the best due to genie-aided CSI knowledge. Although, \ac{5G-NR} LMMSE-Rx performs well with two \ac{DMRS} up to medium Doppler, its performance degrades significantly for higher Doppler, especially with single \ac{DMRS}.

Further, YHP-NeuralRx \cite{Honkala2021} outperforms the proposed YHP-NeuralRx by up to 0.2 dB across all scenarios, except in the high-Doppler single-DMRS case where the gap increases to $0.5$~dB at $10\%$ BLER. However, this performance gain comes at the cost of a $3.7\times$ increase in energy consumption (see Table \ref{tab:efficiency}). In contrast, the proposed Y-NeuralRx outperforms Y-NeuralRx \cite{Faycal2022} by $0.2$–$1$ dB at $10\%$ BLER while simultaneously reducing the energy consumption by $1.8\times$ (see Table \ref{tab:efficiency}). 

For traditional \ac{NeuralRx} (Table~\ref{tab:arch_resnet_t}) with only \ac{ResNet-T} with two \ac{DMRS}, the YHP-NeuralRx improves the accuracy up to $0.5$~dB at $10\%$ \ac{BLER} compared to Y-NeuralRx due to additional \ac{LS} channel estimation and \ac{DMRS} inputs. While using single \ac{DMRS}, the performance is either the same or slightly worse for higher Doppler since single \ac{DMRS} is not sufficient to deal with the effects of high Doppler. In contrast, the proposed \ac{NeuralRx} (Table~\ref{tab:arch_resnet_ss}) with \ac{ResNet-SS} improves the performance up to $0.25$~dB for two \ac{DMRS} and $1.2$~dB for single \ac{DMRS} at $10\%$ \ac{BLER}. 
This is because single \ac{DMRS} has more scope of performance improvement because channel estimation becomes challenging with increasing Delay/Doppler, and the proposed solution with channel split, random shuffle and \ac{GELU} activation operations strengthens the feature extraction in \ac{NeuralRx}. 

\begin{figure}[t!]
    \centering
    \includegraphics[scale=0.65]{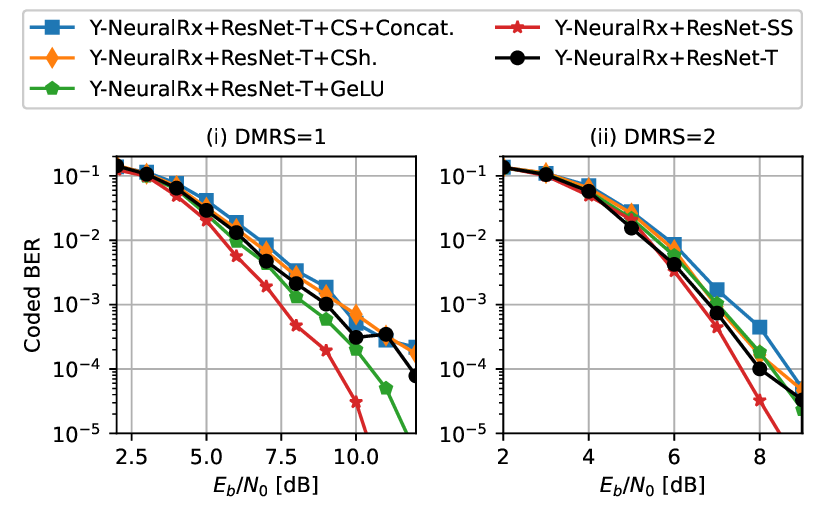}
    \caption{Impact of different components in ResNet-SS.}
   \label{fig:each_layer_eval}
\end{figure}

\subsection{Impact of Components and Computational Cost}
\begin{table}[t]
\caption{Complexity with $C=128$, $K_l=3$ and $D_l=1$.}
\label{tab:efficiency}
\centering 
\begin{tabular}{m{3cm}|>{\raggedleft\arraybackslash}m{1cm}|>{\raggedleft\arraybackslash}m{1.8cm}|>{\raggedleft\arraybackslash}m{1cm}} 
\hline
\textbf{Algorithms} & \textbf{$\text{NParam.}$} & \textbf{\ac{NFLOPs}} & \textbf{Energy} \\ 
\hline
\hline
\textit{5G-NR LS-Rx} &  $0$  & $1,218,540$ & $0.0056$~mJ \\\hline
\textit{5G-NR LMMSE-Rx} &  $0$  &  $3,411,326,635$ & $15.69$~mJ  \\\hline
\hline
\textit{Trad. ResNet-T block} &  $36,096$  &  $259,194,880$  & $1.19$~mJ \\\hline
\textit{Prop. ResNet-SS block} & $9,856$   &  $70,647,808$ & $0.33$~mJ \\\hline
\hline
\textit{Trad. NeuralRx (Table~\ref{tab:arch_resnet_t})} & $264,326$ &  $1,897,419,776$  & $8.73$~mJ   \\\hline
\textit{Prop. NeuralRx (Table~\ref{tab:arch_resnet_ss})} & $198,790$ &  $1,425,822,720$ & $6.56$~mJ  \\\hline
[1] & $604,228$ & $4,996,052,992$ & $24.48$~mJ \\\hline
[2] & $185,988$ & $2,385,295,360$ & $11.68$~mJ \\\hline
\end{tabular}
\end{table}
In Fig.~\ref{fig:each_layer_eval}, we evaluate the impact of channel split (CS), channel concatenate (Concat.), channel shuffle (CSh.) and GeLU activation for different DMRS at for 5M TTI training set. The full ResNet-SS (channel shuffle, split, GeLU, concat) consistently outperforms the traditional ResNet-T and all partial variants that offer little or no improvement, showing that the gain comes from the combined use of all components.

Considering $45$~nm CMOS, the $32$-bit floating-point multiplication and addition need $3.7$~pJ and $0.9$~pJ energy, respectively, totaling to $4.6$~pJ energy for single multiply-and-accumulate operation~\cite{Horowitz2014}. In Table~\ref{tab:efficiency}, we summarize the computational cost for Y-NeuralRx, while YHP-NeuralRx requires additional $9,216$ parameters and $66,060,288$ \ac{FLOPs}, increasing energy-consumption by $3-4\%$ due to additional input channels with $\mathbf{H,P}$. \ac{ResNet-SS} block reduces the energy-consumption by $3.6$ times compared to \ac{ResNet-T} block due to channel split. We can see that \ac{LMMSE} channel estimation requires highest energy-consumption due to a high number of matrix multiplications and matrix inversion while \ac{LS} requires the least (only element-wise division). Overall, proposed \ac{NeuralRx} reduces the energy-consumption by $25\%$ compared to traditional \ac{NeuralRx} due to the combination of \ac{ResNet-T} and \ac{ResNet-SS} blocks. Further, the energy consumption is reduced by a factor of $3.7\times$ and $1.8\times$ compared to~\cite{Honkala2021} and \cite{Faycal2022}, respectively.

\section{Conclusion}
In this work, we designed energy-efficient \ac{ResNet} blocks for the \ac{NeuralRx} by focusing on the number of parameters, \ac{FLOPs}, and \ac{MAC}. We introduce energy-efficient design principles for \ac{ResNet} blocks, that includes utilizing $3\times 3$ kernel size, $1\times 1$ dilation rates and same filter sizes in \ac{ResNet} blocks, to reduce the complexity quadratically compared to \ac{ResNet-T} blocks. We propose a novel \ac{ResNet-SS} block introducing channel-split, channel shuffle and removing element-wise addition, that reduces complexity by $3.6$ times compared to traditional \ac{ResNet-T} block. 
Using \ac{ResNet-T} and \ac{ResNet-SS} blocks, we propose a novel \ac{NeuralRx} architecture design that reduces complexity by $25\%$ compared to existing designs, while improving the training convergence and decoding performance up to $1$~dB. Extensions to multi-layer and DMRS-free/pilotless scenarios are left for future works. 

\bibliographystyle{IEEEtran}
\bibliography{main}

@ARTICLE{Honkala2021,
  author={Honkala, Mikko and Korpi, Dani and Huttunen, Janne M. J.},
  journal={IEEE Transactions on Wireless Communications}, 
  title={DeepRx: Fully Convolutional Deep Learning Receiver}, 
  year={2021},
  volume={20},
  number={6},
  pages={3925-3940},
  doi={10.1109/TWC.2021.3054520}}

@INPROCEEDINGS{Gupta2023,
  author={Gupta et. al., Ankit},
  booktitle={GLOBECOM 2023 - 2023 IEEE Global Communications Conference}, 
  title={Deep Learning-Based Receiver Design for IoT Multi-User Uplink 5G-NR System}, 
  year={2023},
  volume={},
  number={},
  pages={4110-4115},
  doi={10.1109/GLOBECOM54140.2023.10437776}}

@ARTICLE{Faycal2022,
  author={Ait Aoudia, Fayçal and Hoydis, Jakob},
  journal={IEEE Transactions on Wireless Communications}, 
  title={End-to-End Learning for OFDM: From Neural Receivers to Pilotless Communication}, 
  year={2022},
  volume={21},
  number={2},
  pages={1049-1063},
  doi={10.1109/TWC.2021.3101364}}

@ARTICLE{Pihlajasalo2023,
  author={Pihlajasalo et. al., Jaakko},
  journal={IEEE Transactions on Wireless Communications}, 
  title={Deep Learning OFDM Receivers for Improved Power Efficiency and Coverage}, 
  year={2023},
  volume={22},
  number={8},
  pages={5518-5535},
  doi={10.1109/TWC.2023.3235059}}

@ARTICLE{Raviv2023,
  author={Raviv, Tomer and Park, Sangwoo and Simeone, Osvaldo and Eldar, Yonina C. and Shlezinger, Nir},
  journal={IEEE Transactions on Wireless Communications}, 
  title={Online Meta-Learning for Hybrid Model-Based Deep Receivers}, 
  year={2023},
  volume={22},
  number={10},
  pages={6415-6431},
  doi={10.1109/TWC.2023.3241841}}

@ARTICLE{Xie2024,
  author={Xie, Yihang and Teh, Kah Chan and Kot, Alex C.},
  journal={IEEE Transactions on Communications}, 
  title={Comm-Transformer: A Robust Deep Learning-Based Receiver for OFDM System Under TDL Channel}, 
  year={2024},
  volume={72},
  number={4},
  pages={2014-2026},
  doi={10.1109/TCOMM.2023.3343787}}

@ARTICLE{Mei2024,
  author={Mei, Ruru and Wang, Zhugang and Chen, Xuan},
  journal={IEEE Transactions on Cognitive Communications and Networking}, 
  title={CRNN-ResNet: Combined CRNN and ResNet Networks for OFDM Receivers}, 
  year={2024},
  volume={},
  number={},
  pages={1-1},
  doi={10.1109/TCCN.2024.3378225}}

@INPROCEEDINGS{Fischer2022,
  author={Fischer, Moritz Benedikt and Dörner, Sebastian and Krieg, Felix and Cammerer, Sebastian and Brink, Stephan ten},
  booktitle={2022 56th Asilomar Conference on Signals, Systems, and Computers}, 
  title={Adaptive NN-based OFDM Receivers: Computational Complexity vs. Achievable Performance}, 
  year={2022},
  volume={},
  number={},
  pages={194-199},
  doi={10.1109/IEEECONF56349.2022.10051898}}

@INPROCEEDINGS{Korpi2023,
  author={Korpi, Dani and Honkala, Mikko and Huttunen, Janne M.J.},
  booktitle={2023 57th Asilomar Conference on Signals, Systems, and Computers}, 
  title={Deep Learning-Based Pilotless Spatial Multiplexing}, 
  year={2023},
  volume={},
  number={},
  pages={1025-1029},
  doi={10.1109/IEEECONF59524.2023.10477093}}

@INPROCEEDINGS{Horowitz2014,
  author={Horowitz, Mark},
  booktitle={2014 IEEE International Solid-State Circuits Conference Digest of Technical Papers (ISSCC)}, 
  title={1.1 Computing's energy problem (and what we can do about it)}, 
  year={2014},
  volume={},
  number={},
  pages={10-14},
  doi={10.1109/ISSCC.2014.6757323}}

@misc{MobileNet2017,
      title={MobileNets: Efficient Convolutional Neural Networks for Mobile Vision Applications}, 
      author={Andrew G. Howard and Menglong Zhu and Bo Chen and Dmitry Kalenichenko and Weijun Wang and Tobias Weyand and Marco Andreetto and Hartwig Adam},
      year={2017},
      eprint={1704.04861},
      archivePrefix={arXiv},
      primaryClass={cs.CV},
      url={https://arxiv.org/abs/1704.04861}, 
}

@misc{ShuffleNet2017,
      title={ShuffleNet: An Extremely Efficient Convolutional Neural Network for Mobile Devices}, 
      author={Xiangyu Zhang and Xinyu Zhou and Mengxiao Lin and Jian Sun},
      year={2017},
      eprint={1707.01083},
      archivePrefix={arXiv},
      primaryClass={cs.CV},
      url={https://arxiv.org/abs/1707.01083}, 
}

@misc{GELU23,
      title={Gaussian Error Linear Units (GELUs)}, 
      author={Dan Hendrycks and Kevin Gimpel},
      year={2023},
      eprint={1606.08415},
      archivePrefix={arXiv},
      primaryClass={cs.LG},
      url={https://arxiv.org/abs/1606.08415}, 
}

@misc{gupta2024spikingrx,
  title        = {SpikingRx: From Neural to Spiking Receiver},
  author       = {Ankit Gupta and Onur Dizdar and Yun Chen and Stephen Wang},
  howpublished = {arXiv preprint arXiv:2409.05610},
  month        = sep,
  year         = {2024},
  url={https://arxiv.org/abs/2409.05610}, 
}

\end{document}